\documentstyle[12pt]{article}
\def\be{\begin{equation}}
\def\ee{\end{equation}}
\def\bea{\begin{eqnarray}}
\def\eea{\end{eqnarray}}

\renewcommand{\theequation}{\arabic{section}.\arabic{equation}}

\def\case#1/#2{\textstyle\frac{#1}{#2}}
\def\k0{\kappa_{0}}

\title{On Applications of Campbell's Embedding Theorem}
\vspace{0.6cm}
\normalsize
\vspace{2cm}
\author{James E. Lidsey$^{1a}$, Carlos Romero$^{2b}$, Reza Tavakol$^{1c}$\\
\& Steve Rippl$^1$\thanks{e-mail: (a) jel@maths.qmw.ac.uk, (b)
cromero@dfjp.ufpb.br, (c) reza@maths.qmw.ac.uk} \\
\\
$^1$ School of Mathematical Sciences\\
Queen Mary \& Westfield College\\
Mile End Road\\London E1 4NS, UK\\
\\
$^2$ Departamento de F\'{\i}sica\\
Universidade Federal da Para\'{\i}ba\\
C. Postal 5008 - J. Pessoa -Pb\\
58059-970 - Brazil}
\vspace{2cm}
\begin{document}
\sloppy
\maketitle
 
\begin{abstract}
A little known theorem due to Campbell \cite{campbell}
is employed to establish the local embedding of a wide class of 
4--dimensional spacetimes in 
5--dimensional Ricci--flat spaces.
An embedding for the 
class of $n$--dimensional Einstein spaces is also found. 
The local nature of Campbell's theorem 
is highlighted by studying the embedding 
of some lower--dimensional spaces. 
\end{abstract}


\section{Introduction}

\setcounter{equation}{0}

\def\theequation{\thesection.\arabic{equation}}

There has been considerable interest in 
recent years in theories of gravity that contain a different 
number of spatial dimensions from the usual three in  
general relativity (GR). 
One physical  motivation for considering more than
three spatial dimensions arises from the Kaluza--Klein interpretation of the 
fundamental interactions \cite{KK,KKreview,duff,witten}. 
Extra spatial dimensions also 
arise naturally in supergravity \cite{duff} and superstring theories 
\cite{GSW} and 
may have played an important role in the evolution of the 
very early Universe \cite{KKreview,KT}.

A new version of 
5--dimensional GR has recently been developed by Wesson and others
\cite{wesson,wesson1,wesson2}. 
In this approach, the energy density and pressure of the
$(3+1)$--dimensional energy--momentum tensor arise directly from the
extra components of the $(4+1)$--dimensional Einstein tensor,
${^{(5)}}G_{ab}$, where it is assumed that ${^{(5)}}G_{ab}=0$. 
Thus, the physics of $(3+1)$--dimensional cosmologies may be
recovered, in principle, from the geometry of $(4+1)$--dimensional,
vacuum GR \cite{wesson1}.  
Direct calculations have verified that the
spatially flat, perfect fluid cosmologies may be derived in this way
\cite{wesson1,wesson2}.

Lower--dimensional theories of gravity have also
been extensively studied in recent years \cite{2D}. 
These theories are interesting because they may provide a solvable framework 
within which many of the technical and conceptual problems associated with 
quantum gravitational effects in $(3+1)$ dimensions may be addressed. 
A question  that naturally arises in these studies, however, is 
the extent to which the results and intuitions obtained
in lower dimensions may be directly carried 
 over to the $(3+1)$--dimensional environment and vice versa 
\cite{rrt95,verbin}. In general, the 
precise relationship between these theories and $(3+1)$--dimensional 
Einstein gravity is not clear. For example, GR 
does not exhibit a Newtonian limit
in $(2+1)$ dimensions \cite{nolimit}, whereas other theories, 
such as a modified version of 
the Brans--Dicke theory, do have such a limit \cite{verbin}.

An investigation into how lower-- and higher--dimensional theories of
gravity are related to 4--dimensional GR is therefore well
motivated from a physical point of view. A potential bridge between
gravitational theories of different dimensionality may be found by
employing the embedding relationships that exist between spaces and
the main purpose of this paper is to investigate such relationships
further.

The embedding 
of  manifolds in higher dimensions is also 
interesting from a purely mathematical point of view. 
For example, it allows an 
alternative, invariant classification of 
known solutions to Einstein's field 
equations to be made \cite{exact}. Furthermore, the embedding 
method may lead to new solutions. 
Indeed, the maximal analytic extension of the 
Schwarzschild solution was independently 
found in this way \cite{fronsdal}.

A number of embedding theorems are in existence.
It is well known that an analytic Lorentzian space $V_n(s,t)$, 
with $s$ spacelike and $t$ timelike 
dimensions, where $n=s+t$, can be locally 
and isometrically embedded into  a higher 
dimensional, pseudo--euclidean  space $E_N(S,T)$, 
where $N=S+T$, $n \le N \le n(n+1)/2$, and $S \ge s$ and $T \ge t$ are 
positive integers \cite{friedman}.
The line 
element of $E_N$ is given by $ds^2=e_A (dx^A)^2$, where $A= (0, 1, \ldots
, N-1)$ and $e_A =\pm 1$.  Thus, no more than ten 
dimensions are required to embed all 4--dimensional 
solutions to Einstein's field equations. On the other 
hand, if the Ricci tensor of $V_n(s,t)$ is zero, then 
$N\ge n+2$ \cite{kasner}. This implies that 
no curved, 4--dimensional solutions to the vacuum Einstein equations 
can be locally embedded in a 5--dimensional flat space; the minimal 
embedding space is $E_6$ \cite{sch}.

There also exists a theorem due to 
Campbell \cite{campbell,magaard}:

\vspace{.1in}

{\bf Theorem}: {\em Any analytic Riemannian space $V_n(s,t)$ can 
be locally embedded in a Ricci--flat, 
Riemannian space $V_{n+1}(\tilde{s}, \tilde{t})$, where 
$\tilde{s}=s$ and $\tilde{t}=t+1$, or $\tilde{s}=s+1$ and $\tilde{t}=t$.}

\vspace{.1in}

Campbell's theorem implies that all solutions to the $n$--dimensional 
Einstein field equations with arbitrary energy--momentum tensor 
can be embedded, at least locally, in a spacetime that is 
itself a solution to $(n+1)$---dimensional, 
vacuum GR \cite{rtz}. (We refer to a local 
embedding in the usual differential
geometric sense, i.e.,  without any direct reference to the global 
topology of the embedding or the embedded spaces). 
This theorem is 
therefore closely related to Wesson's procedure 
\cite{wesson},  since it implies that 
any  4--dimensional cosmology
may be locally embedded in a 5--dimensional 
Ricci--flat space, at least in principle.
Very little reference to Campbell's theorem can be found in the literature
(see, however, Magaard \cite{magaard} and Goenner \cite{goenner}).
Recently, 
Romero, Tavakol and Zalaletdinov \cite{rtz} outlined the proof of this 
theorem in a modern notation and discussed its relationship 
with Wesson's procedure \cite{wesson}. They also emphasised its constructive
nature with the help of some concrete examples. 

Here we 
investigate some further applications
of Campbell's theorem. 
Section 2 summarizes the main 
points of the theorem.
In Sections 3 and 4 we consider the 
embedding of the general class of 4--dimensional 
spacetimes that admit a non--twisting null Killing vector \cite{exact}. 
In Section 5 we find an embedding space for the 
general class of $n$--dimensional Einstein spaces. 
We then proceed in Section 6 to discuss some  
aspects of Campbell's theorem regarding the local and global embeddings 
of lower--dimensional gravity. We conclude in Section 7.
%
\section{The Embedding Theorem of Campbell}
%

\setcounter{equation}{0}

\def\theequation{\thesection.\arabic{equation}}

We begin by discussing  the embedding 
theorem due to Campbell \cite{campbell,magaard,rtz}.
Consider the  space $V_n(s,t)$ with 
metric ${^{(n)}}g_{\alpha\beta} (x^{\mu})$ and line element\footnote{In this 
paper, Greek indices take values in the range $(0, 1, \ldots ,n-1)$, 
Latin indices run from $(0, 1, \ldots , n)$, semicolons and
commas indicate covariant and partial differentiations respectively
and spacetime metrics have signature $(+,-,-, \ldots)$.}
\be
\label{Dmetric}
{^{(n)}}ds^2 ={^{(n)}}g_{\alpha\beta}(x^{\mu}) dx^{\alpha} dx^{\beta} ,
\ee
and let the local embedding of 
this space in the  manifold  $V_{n+1}(\tilde{s}, \tilde{t})$ 
be given by 
\be
\label{D+1metric} 
{^{(n+1)}}ds^2 =g_{\alpha\beta} (x^{\mu} ,\psi) dx^{\alpha}dx^{\beta} +
\epsilon \phi^2(x^{\mu} ,\psi )d\psi^2 ,
\ee
where $\epsilon =\pm 1$ and $\psi$ is the coordinate that spans the 
extra dimension.
It is assumed that $g_{\alpha\beta}$, when
restricted to a hypersurface $\psi =\psi_0$,
results in $^{(n)}g_{\alpha\beta}$:
\be
g_{\alpha\beta}(x^{\mu} ,\psi_0 )= {^{(n)}}g_{\alpha\beta} (x^{\mu}).
\label{init}
\ee

According to Campbell's theorem \cite{campbell}, the
functional form of the higher--dimensional metric coefficients 
(\ref{D+1metric}) can 
be determined if functions 
$\Omega_{\alpha\beta} (x^{\mu} ,\psi )$
may be found which satisfy the set of conditions
\bea
\label{omega1}
\Omega_{\alpha\beta} =\Omega_{\beta\alpha} \\
\label{omega2}
{\Omega^{\alpha}}_{\beta ; \alpha} =\Omega_{,\beta} \\
\label{omega3}
\Omega_{\alpha\beta}\Omega^{\alpha\beta} -\Omega^2 =- 
\epsilon~ {^{(n)}}R 
\eea
on some hypersurface $\psi =\psi_0$  and if the functions 
$g_{\alpha\beta}$ and $\Omega_{\alpha\beta}$ evolve in accordance
with the equations
\be
\label{metricdevelop}
\frac{\partial g_{\alpha\beta}}{\partial \psi} =-2\phi
\Omega_{\alpha\beta},
\ee
\be
\label{omegadevelop}
\frac{\partial {\Omega^{\alpha}}_{\beta}}{\partial \psi} =\phi
\left( -\epsilon {^{(n)}}{R^{\alpha}}_{\beta}
+\Omega {\Omega^{\alpha}}_{\beta}
\right) +\epsilon g^{\alpha\lambda}  \phi_{; \lambda\beta} ,
\ee
respectively\footnote{We note here
that 
with $\phi=1,~~n=3$,
there is a clear parallel with the language used in the 
$1+3$ decomposition employed in the ADM formalism \cite{adm}
and the initial value formulation of GR \cite{init}.
This can be seen through the following identifications:
$$t\rightarrow \psi,~~~ h_{\alpha\beta}\rightarrow ~ {^{(3)}}g_{\alpha\beta},
~~K_{\alpha\beta}\rightarrow \Omega_{\alpha\beta} ,$$
where $h_{\alpha\beta}$ is the metric of the 3-space
and $K_{\alpha\beta}$ is the extrinsic curvature.}. 
In these expressions, 
${\Omega^{\alpha}}_{\beta} \equiv {^{(n)}}g^{\alpha\lambda}
\Omega_{\lambda\beta}$, $\Omega \equiv {^{(n)}}g^{\alpha\beta}
\Omega_{\alpha\beta}$ and ${^{(n)}}R \equiv
{^{(n)}}R_{\mu\nu} {^{(n)}}g^{\mu\nu}$. 

If Eqs. (\ref{metricdevelop}) and (\ref{omegadevelop}) are evaluated
on the hypersurface $\psi =\psi_0$, it can be shown that
Eqs. (\ref{omega1})--(\ref{omegadevelop}) are equivalent to the
vacuum, $(n+1)$--dimensional GR field equations
${^{(n+1)}}R_{ab}(x^{\mu}, \psi_0)=0$ \cite{rtz}. Moreover, it can be
proved that Eqs. (\ref{omega1})--(\ref{omega3}) are valid for all
$\psi$ in the neighbourhood of $\psi_0$ when
Eqs. (\ref{metricdevelop}) and (\ref{omegadevelop}) are satisfied
\cite{campbell}. This implies that the Ricci tensor of $V_{n+1}$
vanishes for any $\psi$ in the neighbourhood of $\psi_0$.
Consequently, Eq. (\ref{D+1metric}) may be viewed as an embedding of
the metric (\ref{Dmetric}) in a Ricci--flat, $(n+1)$--dimensional
space.

Applications of Campbell's theorem considered in this paper are based
up on the integrability of Eq. (\ref{omegadevelop}).  We consider the
embedding of spaces with vanishing Ricci scalar curvature
$({^{(n)}}R=0)$ and also find an embedding 
for the class of Einstein spaces $({^{(n)}} R={\rm
constant})$. In the former case, one solution to
Eqs. (\ref{omega1})--(\ref{omega3}) is given by $\Omega_{\alpha \beta}
=0$. It should be emphasized, however, that this does not necessarily
represent the most general solution possible. Thus, the embedding of
spaces with ${^{(n)}}R=0$ may be divided into two subclasses. These
correspond to embeddings where all the components of
$\Omega_{\alpha\beta}$ vanish and to those where some (or all) of the
components are non--trivial.

As an example of an application of this theorem, 
we will conclude this Section by discussing the subclass of embeddings where 
$\Omega_{\alpha\beta}=0$.  
It follows from Eq. (\ref{metricdevelop}) that $\partial g_{\alpha 
\beta} /\partial \psi =0$ and 
$g_{\alpha\beta}$ is therefore independent of the extra coordinate
$\psi$.  This implies that $g_{\alpha\beta} = {^{(n)}}g_{\alpha\beta}$
in the neighbourhood of the hypersurface $\psi =\psi_0$.  The one
remaining equation that needs to be solved is Eq. (\ref{omegadevelop})
which simplifies to
\be
\label{phidevelop}
{^{(n)}} g^{\alpha\lambda} \phi_{; \lambda \beta} = 
{^{(n)}}{R^{\alpha}}_{\beta} \phi .
\ee
We conclude, therefore, 
that the embedding metric is given by
\be
\label{flatflat}
{^{(n+1)}}ds^2 = {^{(n)}}g_{\alpha\beta} dx^{\alpha} dx^{\beta} +\epsilon
\phi^2 d\psi^2,
\ee
where $\phi$ is a solution to Eq. (\ref{phidevelop}). 

Taking the trace of Eq. (\ref{phidevelop})  implies that $\phi$ must
satisfy the massless Klein--Gordon equation, ${^{(n)}}g^{\alpha\beta}
\phi_{;\alpha\beta} =0$. If we assume an embedding of this form, 
therefore, a necessary, but not sufficient, condition on $\phi$ is that 
it be an harmonic function of $x^{\mu}$. 
This restriction often provides valuable insight into 
the generic form that $\phi$ must take if it is to satisfy 
the full set of differential equations (\ref{phidevelop}). We remark that
similar conclusions hold when $V_n$ is Ricci--flat. Indeed, 
one solution to Eq. (\ref{phidevelop}) in this  case is  $\phi =1$ 
and this provides a simple proof of theorem III 
of Romero {\em et al.}  \cite{rtz}. 

An interesting consequence of this 
embedding is that it may 
be repeated indefinitely, at least in principle. That is, the Ricci--flat 
space with metric (\ref{flatflat}) may itself be 
embedded in an $(n+2)$--dimensional, Ricci--flat space with a
line--element given by 
${^{(n+2)}}ds^2 = {^{(n+1)}}ds^2 + \varphi^2 d\theta^2$, where 
$\theta$ represents the extra coordinate and $\varphi =\varphi ( x^{\mu}
, \psi , \theta)$ is an harmonic function satisfying 
${^{(n+1)}}g^{\alpha\lambda} \varphi_{; \lambda \beta} = 0$. 
Thus, once a given Lorentzian  space $V_n$ has been embedded in 
a Ricci--flat space $V_{n+1}$, further embeddings 
in Ricci--flat spaces of  progressively higher dimensions can be considered. 

Having summarized the steps that need to be 
taken when applying Campbell's theorem, 
we proceed in the following Sections 
to investigate the local embedding of a 
wide class of Lorentzian spaces 
with one timelike dimension. We
begin by considering the embedding of 4--dimensional
spacetimes that admit a non--twisting null Killing vector
field ${\bf k}$,
where $k_{(\mu ;\nu)} =0,~~ k_{\mu}k^{\mu} = 0$ 
and $k_{\left[ \mu \right. } k_{
\left. \nu ;\rho \right]} =0$. It can be shown that
there are two classes of metrics that admit a
Killing vector of this form \cite{exact}, depending up on whether
${\bf k}$ is covariantly constant, $k_{\mu ;\nu} =0$, 
or whether it satisfies the less severe restriction $k_{(\mu ;\nu)} =0.$

\section{The Embedding of Spacetimes Admitting a 
Covariantly Constant Null Killing Vector Field}

\setcounter{equation}{0}

\def\theequation{\thesection.\arabic{equation}}

Metrics admitting a covariantly constant, 
null Killing vector field ${\bf k}$ have the form
\be
\label{nullc}
ds^2=dudv +fdu^2 -dx^2 -dy^2  ,
\ee
where $k_{\mu}=\partial_{\mu} u$ and 
$f=f(u,x,y)$ is an arbitrary function that is independent of 
the coordinate $v$ \cite{exact}. The  coordinates $(x,y)$ span the 
spacelike 2--surfaces that are orthogonal to ${\bf k}$ 
and the surfaces $u={\rm constant}$ are null. 
The Ricci scalar for these spacetimes vanishes for arbitrary $f$, whilst the 
Riemann and Ricci tensors are given by $R_{\mu\nu\rho\sigma}= -2k_{\left[ \mu 
\right.}\partial_{ \left. \nu \right]}
\partial_{\left[ \rho \right.}fk_{\left. \sigma \right]}$
and $R_{\mu\nu}=\frac{1}{2} ( \partial^2_Tf ) k_{\mu}k_{\nu}$, respectively, 
where $\partial^2_T$ is the Laplacian on the transverse 2--surfaces. 
The only non--zero components of these tensors are 
\be
\label{rie}
R_{uxux} =-\frac{1}{2} f_{,xx}, \quad 
R_{uxuy} =-\frac{1}{2} f_{,xy}, \quad 
R_{uyuy} =-\frac{1}{2} f_{,yy}
\ee
and 
\be
\label{ricci}
R_{uu} = \frac{1}{2} \left( f_{,xx} +f_{,yy} \right)  
\ee
and it follows from Eq. (\ref{rie}) that linear terms in $f$ of the form 
$a(u) +b_i(u)x^i$ do not affect the Riemann tensor. They 
can therefore be transformed away. 

The class of spacetimes given by Eq. (\ref{nullc}) 
is physically very interesting. They 
are known as {\em plane waves} when $f(u,x^i) = h_{ij}(u)x^ix^j$ for some 
symmetric function $h_{ij} (u)$. 
These solutions with $h_{ii}=0$ were first discussed by Brinkman 
\cite{brinkman1923}.
A purely gravitational wave is characterized by the condition 
$h_{ii}(u)=0$ and a purely electromagnetic wave corresponds to 
$h_{ij}(u) =h(u)\delta_{ij}$, where $h(u) \ge 0$ \cite{penrose}. 
The amplitudes 
of the gravitational and electromagnetic waves are 
given by the trace--free part of $h_{ij}$ and by $[{\rm Tr}(h_{ij})]^{1/2}$, 
respectively. In general, the amplitudes may be arbitrary functions of $u$.

When $f$ is a solution to the Laplace equation $\partial_T^2 f=0$, 
the manifold is Ricci--flat. In this case, 
Eq. (\ref{nullc}) 
represents the most general, 4--dimensional
 solution to the vacuum Einstein field equations 
with a covariantly constant null vector \cite{exact}. 
The dependence of $f$ on $u$
may be arbitrary and many different solutions can 
therefore be considered. More general solutions to Laplace's equation, 
where $h_{ij}$ also depends on $x^k$, are known as {\em plane--fronted 
waves}. 

Plane--fronted waves 
are solutions to any gravitational  theory whose 
field equations are given in terms of a second--rank tensor 
derived from the curvature tensor and its derivatives \cite{HS}.
This property may be traced to the fact that the curvature is null. 
Included in this class of theories is string theory \cite{perry}. 
Indeed, plane--fronted waves are {\em exact} solutions to 
the classical equations of motion to {\em all} orders in $\sigma$--model 
perturbation theory \cite{exactstring,HS}. 
Exact solutions that include non--trivial dilaton and antisymmetric tensor 
fields can also be found and correspond to the case where $\partial^2_Tf$ 
is an arbitrary function of $u$ \cite{HS}. 

Since $R=0$ for all $f$, 
we may begin by choosing $\Omega_{\alpha\beta}=0$. An
embedding metric is therefore 
given by Eq. (\ref{flatflat}),  where $\phi=\phi(x^{\mu} ,
\psi)$ satisfies Eq. (\ref{phidevelop}).
This equation represents the set of coupled 
differential equations: 
\bea
\label{c1}
f_{,x}\phi_{,v} -\phi_{,ux} =0 \\
\label{c2}
f_{,y}\phi_{,v} - \phi_{,uy} =0 \\
\label{c3}
-2f_{,u}\phi_{,v} -f_{,x}\phi_{,x}-f_{,y}\phi_{,y} +2 \phi_{,uu}
= \left( f_{,xx} +f_{,yy} \right) \phi \\
\label{c4}
\phi_{,uv}=\phi_{,vv}=\phi_{,vx}=\phi_{,vy}=
\phi_{,xx}=\phi_{,xy}=\phi_{,yy}=0.
\eea
Differentiating Eqs. (\ref{c1}) and (\ref{c2}) both 
with respect to $x$ and $y$ implies that 
\be
\label{nov}
f_{,xx}\phi_{,v}=f_{,xy}\phi_{,v}=f_{,yy}\phi_{,v}=0,
\ee
where we have employed Eq. (\ref{c4}). It follows from 
Eqs. (\ref{rie}) and (\ref{nov}) that the Riemann curvature tensor 
must vanish 
if  $\phi_{,v} \ne 0$, thereby implying that the spacetime is flat. 
Consequently, $\phi$ must be 
independent of $v$ when  $\Omega_{\alpha\beta}=0$. In this case, 
Eqs. (\ref{c1})--(\ref{c3}) simplify to
\bea
\label{d1}
\phi_{,ux} =0 \\
\label{d2}
\phi_{,uy}=0 \\
\label{d3}
-f_{,x}\phi_{,x}-f_{,y}\phi_{,y} +2 \phi_{,uu}
= \left( f_{,xx} +f_{,yy} \right) \phi  .
\eea

The embedding metric 
is determined once Eqs. (\ref{d1})--(\ref{d3}) have been solved
subject to the constraints (\ref{c4}). 
We will now consider the vacuum and non--vacuum cases in turn. 
\subsection{The Embedding of Vacuum Plane--fronted Waves}
The right--hand side of Eq. (\ref{d3}) vanishes 
for the vacuum solutions. 
Differentiating this equation  with respect to both $x$ and $y$ then 
implies that 
\bea
f_{,xx}\phi_{,x} +f_{,xy}\phi_{,y} =0 \nonumber \\
f_{,xy}\phi_{,x} +f_{,yy}\phi_{,y} =0 
\eea
and combining these two equations implies that
\bea
\label{fconstraint}
f_{,xx} \left( \phi_{,x}^2 +\phi^2_{,y} \right) =0 \nonumber \\
f_{,xy} \left( \phi_{,x}^2 +\phi^2_{,y} \right) =0  .
\eea
If the embedded spacetime is vacuum and has non-zero curvature,
the second derivatives of $f$ with respect to $x$ and $y$ must be 
non--vanishing. Consequently, Eq. (\ref{fconstraint}) 
can only be satisfied if $\phi^2_{,x}=-\phi^2_{,y}$. However, 
$\phi$ should be a real function if the embedding 
spacetime is to be physical. 
Thus, $\phi$ must be independent of both $x$ and $y$. 

The only non--trivial constraint that remains in Eqs. (\ref{d1})--(\ref{d3}), 
therefore, is that 
$\phi_{,uu}=0$ and this has the general solution $\phi= a(\psi) +b(\psi) u$, 
where $a$ and $b$ are arbitrary functions of the fifth coordinate $\psi$. 
One possible embedding of 4--dimensional, 
vacuum, plane--fronted waves in a 5--dimensional, Ricci--flat 
manifold is therefore given by 
\be
\label{a}
ds^2=dudv +fdu^2 -dx^2 -dy^2 -\left( a(\psi) +b(\psi)u \right)^2  
d\psi^2  .
\ee

A second local embedding of these plane--fronted waves
can be found by assuming that 
\bea
\label{ansatz}
{\Omega_{\alpha\beta}} = 
\left\{ \begin{array}{ll}
        f/(2\psi_0 )    & \mbox{if $\alpha =\beta =u$} \\
        0               & \mbox{otherwise}
 	\end{array} \right. 
\eea
on the hypersurface $\psi = \psi_0$. In this case, the only 
non--zero components of ${\Omega^{\alpha}}_{\beta}$ and $\Omega^{\alpha\beta}$
are ${\Omega^v}_u = 2\Omega_{uu}$ and $\Omega^{vv}=4\Omega_{uu}$, 
respectively, where indices have been raised with 
${^{(4)}}g^{\alpha\beta}$. It follows immediately that $\Omega =
\Omega_{\alpha\beta}\Omega^{\alpha\beta} =0$, so Eq. (\ref{omega3}) 
is satisfied. Moreover, Eq. (\ref{omega2}) simplifies to 
${\Omega^v}_{u;v} =0$. Since both ${\Omega^{\alpha}}_{\beta}$ 
and ${^{(4)}}g_{\alpha\beta}$ 
are independent of $v$, however, this condition is also satisfied. 

Eq. (\ref{omegadevelop}) is solved by specifying $\phi =-1$, 
since ${\Omega^{\alpha}}_{\beta}$ is  independent of $\psi$. Thus, 
the solution to Eq. (\ref{metricdevelop}) that satisfies the initial 
conditions (\ref{init}) on the hypersurface $\psi=\psi_0$ is given by 
\bea
g_{\alpha\beta} = 
\left\{ \begin{array}{ll}
         (\psi /\psi_0 ) f                & \mbox{if $\alpha =\beta =u$} \\
        {^{(4)}}g_{\alpha\beta}          & \mbox{otherwise.}
 	\end{array} \right. 
\eea
It follows, therefore, that when indices are raised with 
$g^{\alpha\beta}$, the only non--zero components of 
${\Omega^{\alpha}}_{\beta}$ and $\Omega^{\alpha\beta}$ are 
 ${\Omega^v}_u$ and  $\Omega^{vv}$, as before. Thus, Eqs. (\ref{omega1}) 
and (\ref{omega3}) are valid for arbitrary $\psi$. The 
same conclusion holds 
for Eq. (\ref{omega2}), since $g^{\alpha\beta}$ is itself 
independent of $v$. 

We may conclude, therefore, that 
the 5--dimensional embedding for this {\em ansatz} is given by 
\be
\label{b}
ds^2= dudv + \frac{\psi}{\psi_0} f du^2 -dx^2 -dy^2 -d\psi^2.
\ee
It may be verified by direct calculation that 
this space is Ricci-flat. 
Its curvature differs from that of Eq. (\ref{a}), however. In particular, 
we find that 
$R_{uxu \psi} =-(1/2\psi_0)f_{,x}$ and $R_{uyu \psi}=-(1/2\psi_0)f_{,y}$, 
whereas 
these components vanish for the spacetime corresponding to Eq. (\ref{a}). 

\subsection{The Embedding of Electromagnetic Waves and Exact String 
Backgrounds}
We may consider the more general class of spacetimes characterized  by 
\be
\label{stringf}
f (u , x^k)=h_{ij} (u) x^i x^j +f_T(u, x^k)  ,
\ee
where $f_T$ is an 
arbitrary solution to $\partial_T^2 f_T=0$. These spacetimes are 
not Ricci--flat if ${\rm Tr}(h_{ij}) \ne 0$, since 
$\partial_T^2 f = 2(h_{11}+h_{22})$. 

Now, the general form of $\phi$ consistent with Eqs. (\ref{c4}), (\ref{d1}) 
and (\ref{d2}) is given by 
\be
\label{genphi}
\phi =a(u) +bx+cy  ,
\ee
where $a(u)$ is an arbitrary function 
of $u$ and $b$ and $c$ are arbitrary constants. (We assume for 
simplicity that $\phi$ is independent of the fifth coordinate). The 
embedding spacetime may therefore be determined by finding 
a solution to Eq. (\ref{d3}) that is consistent with 
Eq. (\ref{genphi}). 

Let us begin with the simpler case where $b=c=0$, so that 
$\phi$ is a function only of $u$. 
Substitution of Eqs. (\ref{stringf}) and (\ref{genphi}) 
into Eq. (\ref{d3}) then implies that
\be
\label{e1}
\frac{d^2 a}{du^2} =\left( h_{11}+h_{22} \right) a  .
\ee
In this case, the embedding metric is given by
\be 
ds^2=dudv + fdu^2 -dx^2 -dy^2 -a^2 (u) d\psi^2
\ee
and may be expressed  in a closed form whenever an exact solution 
to Eq. (\ref{e1}) can be found for a given $h_{ij}(u)$. 
This embedding is general, in the sense that the amplitude 
$h_{ij}$ is an arbitrary function of $u$. The functional
form of $\phi$ is also independent of $f_T$, so we may  consider 
arbitrary forms for this latter function, 
subject to the condition that it satisfies 
the Laplace equation $\partial^2_Tf_T=0$. 
Eq. (\ref{equation}) corresponds to the embedding 
of an electromagnetic plane wave of arbitrary amplitude when 
 $h_{12} = f_T=0$ and $h_{11}=h_{22}$. 
In this case, a space with vanishing Weyl tensor is embedded 
in a space with vanishing Ricci tensor of the form
\be
\label{equation}
ds^2=dudv +\left( \frac{1}{2a}\frac{d^2a}{du^2} \right) \left( 
x^2 +y^2 \right) du^2 -dx^2 -dy^2 -a^2(u) d\psi^2  .
\ee

An embedding is also possible if $(b,c) \ne 0$ and $f_T=0$.  
In this case, Eq. (\ref{e1}) still applies, but the components 
of $h_{ij}$ are restricted by the additional constraints
\bea
\label{e2}
\left( 2h_{22}+h_{11} \right)^{1/2} \left( 2h_{11}+h_{22} \right)^{1/2}
=\mp h_{12} \nonumber \\
b =\pm \left( \frac{2h_{22}+h_{11}}{2h_{11}+h_{22}} \right)^{1/2} c  .
\eea
We find that the embedding metric is given by
\be
ds^2=dudv +h_{ij}x^ix^j du^2 -dx^2 -dy^2 - \left[ a(u) +bx +cy \right]^2 
d\psi^2
\ee
when Eqs. (\ref{e1}) and (\ref{e2}) are satisfied. 

\section{The Embedding for Spacetimes Admitting a 
Non--Constant Null Killing Vector Field}

\setcounter{equation}{0}

\def\theequation{\thesection.\arabic{equation}}

If the null Killing vector is not 
(covariantly) constant, the metric is given by 
\be
\label{stockum}
ds^2=2xdu(dv +mdu) -x^{-1/2} (dx^2+dy^2)  ,
\ee
where $m=m(u,x,y)$ is an arbitrary function and is independent 
of $v$ \cite{exact}. The Ricci curvature scalar of these spacetimes
vanishes for arbitrary $m$ and 
the only non--trivial component of the Ricci tensor is given by
\be
R_{uu} = x^{1/2} \left( \left( xm_{,x} \right)_{,x} +xm_{,yy} \right).
\ee
We will consider the subset of spacetimes that are solutions to vacuum
GR. This includes a wide class of spaces, since the dependence of $m$
on $u$ is arbitrary. When $m$ is independent of $u$ and
$R_{uu}=0$ the spacetimes (\ref{stockum}) are
the stationary van Stockum solutions \cite{exact}.

Following the discussion of Section 2, 
we choose $\Omega_{\alpha\beta} =0$, since the Ricci scalar is zero. The  
$(u,u)$, $(u,y)$, 
$(x,u)$, $(x,v)$, $(x,y)$ and $(y,y)$ components of Eq. (\ref{phidevelop})
are then given by 
\bea
\label{f1}
x^{1/2} \phi_{,x} -2 \phi_{,uv} =0 \\
\label{f2}
\phi_{,vy} =0 \\
\label{f3}
2x m_{,x}\phi_{,v} -2x \phi_{,ux} +\phi_{,u} =0 \\
\label{f4}
2x \phi_{,vx}-\phi_{,v}=0 \\
\label{f6}
4x\phi_{,xy} +\phi_{,y} =0 \\
\label{f7}
4x\phi_{,yy}-\phi_{,x} =0,  
\eea
respectively. However, differentiation of Eq. (\ref{f7}) 
with respect to $v$ implies that $\phi_{,vx}=0$, 
where we have employed Eq. (\ref{f2}). Thus, Eq. (\ref{f4}) 
implies that $\phi$ must be  independent of $v$, but it then 
follows from Eq. (\ref{f1}) 
that $\phi$ must also be independent of $x$. 
Eq. (\ref{f6}) then implies that $\phi$ 
must also be independent of $y$ and, finally, Eq. (\ref{f3}) 
implies that $\phi$ is independent of $u$. 
In conclusion, therefore, the 
only consistent solution is that $\phi=\phi (\psi)$ when 
$\Omega_{\alpha\beta} =0$. 

An alternative embedding may be found by assuming the {\em ansatz}
\bea
\label{ansatz1}
{\Omega_{\alpha\beta}} = 
\left\{ \begin{array}{ll}
        xm/\psi_0   & \mbox{if $\alpha =\beta =u$} \\
        0               & \mbox{otherwise.}
 	\end{array} \right. 
\eea
The argument is similar to that followed in the previous Section. 
Eqs. (\ref{omega1})--(\ref{omega3}) are satisfied on a 
particular hypersurface $\psi =\psi_0$, because 
${^{(4)}}g^{uu} =0$ and ${^{(4)}}g^{\alpha\beta}$ and $\Omega_{\alpha\beta}$ 
are independent of $v$. One solution to Eq. (\ref{omegadevelop}) is 
given by $\phi =-1$ and this implies that the integral of 
Eq. (\ref{metricdevelop}) is given by 
\bea
g_{\alpha\beta} = 
\left\{ \begin{array}{ll}
         2xm(\psi /\psi_0 )                & \mbox{if $\alpha =\beta =u$} \\
        {^{(4)}}g_{\alpha\beta}            & \mbox{otherwise.}
 	\end{array} \right. 
\eea
The embedding metric is therefore given by
\be
ds^2=2xdudv + 2xm \left( \frac{\psi}{\psi_0} \right) du^2 -x^{-1/2} dx^2 
-x^{-1/2}dy^2 - d\psi^2.
\ee
It may be verified that Eqs. (\ref{omega1})--(\ref{omega3}) remain valid 
when indices are raised with $g^{\alpha\beta}$, so they 
are valid for all $\psi$. 

To summarize thus far, we have found  embedding spaces for 
the general class of metrics that admit a non--twisting, 
null Killing vector field. 
All these spacetimes have vanishing curvature scalar, however. We 
therefore extend our analysis in the following Section to 
include the class of spacetimes for which 
the curvature scalar  is covariantly constant.

\section{The Embedding of Einstein spaces}

\setcounter{equation}{0}

\def\theequation{\thesection.\arabic{equation}}

The class of $n$--dimensional Einstein spaces is defined by  
\be
\label{scalar}
{^{(n)}}{R^{\alpha}}_{\beta} = \left( \frac{\kappa}{n} \right)
{^{(n)}}{\delta^{\alpha}}_{\beta} , \qquad {^{(n)}}R =\kappa  ,
\ee
where $\kappa$ is an arbitrary constant \cite{HE}. For $n \ge 3$, we 
may define $\Lambda  \equiv (2-n)\kappa/(2n)$. 
This may be viewed 
as a cosmological constant in the vacuum Einstein field 
equations ${^{(n)}}{G^{\alpha}}_{\beta} = \Lambda 
{^{(n)}}{\delta^{\alpha}}_{\beta}$. 

We proceed by specifying $\phi =1$. Eq. (\ref{omegadevelop}) then reduces to 
\be
\label{44}
\frac{\partial {\Omega^{\alpha}}_{\beta}}{\partial \psi} = 
\frac{- \epsilon\kappa}{n} {\delta^{\alpha}}_{\beta} + \Omega 
{\Omega^{\alpha}}_{\beta}.
\ee
This equation admits the exact solution
\be
\label{hypersolution}
{\Omega^{\alpha}}_{\beta} =a \psi^{-1} {\delta^{\alpha}}_{\beta}
\ee
on the specific hypersurface 
\be 
\label{hyper}
\psi =\psi_0 =\pm \left( \frac{an (1+an)}{\epsilon\kappa} \right)^{1/2},
\ee
where $a$ is a constant. 
When $\epsilon =-1$ (corresponding to an extra spacelike coordinate), 
we require for consistency that $-1/n < a <0$ if $\kappa>0$ and 
$a<-1/n$ or $a>0$ if $\kappa <0$. Conversely, for $\epsilon =+1$, we 
require $a>0$ or $a<-1/n$ for $\kappa >0$ and $-1/n < a <0$ if 
$\kappa <0$. 

We will assume 
for the moment that the solution (\ref{hypersolution}) 
is valid for arbitrary values of $\psi$ 
when ${^{(n)}}{R^{\alpha}}_{\beta}$ is calculated with 
$g_{\alpha\beta} (x^{\mu} ,\psi ) $
rather than with the original metric ${^{(n)}}g_{\alpha\beta} (x^{\mu})$.
(The validity of this assumption will be verified shortly for $a=-1$). 
We may now integrate Eq. (\ref{metricdevelop}) subject to 
the initial conditions (\ref{init}). We find that 
\be
\label{ghyper}
g_{\alpha\beta} (x^{\mu} ,\psi) 
 =  \psi^{-2a} \left( \frac{an (1+an)}{\epsilon\kappa} \right)^a  
 {^{(n)}}g_{\alpha\beta} (x^{\mu}).
\ee
The functions $\Omega_{\alpha\beta}$ are then determined by this equation 
and Eq. (\ref{hypersolution}). The result is
\be
\Omega_{\alpha\beta} = a \psi^{-1-2a} 
\left( \frac{an (1+an)}{\epsilon \kappa} \right)^{a} 
{^{(n)}}g_{\alpha\beta}.
\ee

However, Eqs. (\ref{omega1})---(\ref{omega3}) must also 
be satisfied. The functions $\Omega_{\alpha\beta}$ are symmetric, 
so Eq. (\ref{omega1}) is clearly valid. Moreover, they are independent 
of $x^{\mu}$, so Eq. (\ref{omega2}) is also consistent. On 
the other hand, Eq. (\ref{omega3}) is more restrictive because 
$\Omega = an \psi^{-1} \ne 0$ in general. Indeed, this equation is 
only satisfied if $a=-1$.
Thus, the solution (\ref{hypersolution}) only applies 
if $\epsilon \kappa >0$, which implies that 
$\epsilon = -1$ $(\epsilon =+1 )$ for a positive (negative) $\Lambda$. 

Finally, Eq. (\ref{omegadevelop}) must be considered 
when ${^{(n)}}{R^{\alpha}}_{\beta}$ is 
calculated with the $n$--dimensional part of the 
$(n+1)$--dimensional embedding metric. This will establish the validity
of this embedding procedure.  For a given  metric $g_{\alpha\beta}$, we 
may perform the conformal transformation $\tilde{g}_{\alpha\beta} 
= k g_{\alpha\beta}$, where $k$ is constant. 
It follows that ${\tilde{{R}^{\alpha}}}_{\beta} = k^{-1} 
{R^{\alpha}}_{\beta}$ and $\tilde{R} =k^{-1} R$. 
Using this property, we are able to calculate ${^{(n)}}{R^{\alpha}}_{\beta} $ 
 with the metric $g_{\alpha\beta} (x^{\mu} ,\psi ) = k ^{(n)} 
g_{\alpha\beta}$, since $k = 
\epsilon \kappa \psi^2/[n(n-1) ]$ is 
a constant. We find  that 
\be
\label{newricci}
{^{(n)}}{R^{\alpha}}_{\beta} (x^{\mu} ,\psi ) = 
k^{-1} {R^{\alpha}}_{\beta} (x^{\mu}) =\left[ 
\frac{n-1}{\epsilon \psi^2} 
\right] {^{(n)}}{\delta^{\alpha}}_{\beta}
\ee
and one  can further show by an analogous argument that 
\be
\label{newscalar}
{^{(n)}}R(x^{\mu} ,\psi ) = k^{-1} R =\frac{n(n-1)}{\epsilon \psi^2}  .
\ee
Direct substitution of Eqs. (\ref{newricci}) 
and (\ref{newscalar}) then implies 
that Eqs. (\ref{omega3}) and (\ref{omegadevelop}) 
are valid for arbitrary $\psi$. Consequently, the embedding metric 
is locally Ricci--flat for every value of $\psi$, as required, and it
is given by
\be
\label{einsteinembed}
{^{(n+1)}}ds^2 = \left[ \frac{\epsilon \kappa}{n(n-1)} \psi^2 \right] 
{^{(n)}}g_{\alpha\beta} (x^{\mu} ) dx^{\alpha} dx^{\beta}  +
\epsilon d\psi^2.
\ee

The embeddings that we have considered in this paper thus far 
are local, in the sense that no reference was made to the 
global topology of either the embedded or embedding space. This 
is because Campbell's theorem is a local theorem. In the next 
Section, we shall highlight the local nature of this theorem further 
by investigating the embeddings of some lower--dimensional spacetimes.

\section{Local and Global Embedding of Spaces with 
Lower Dimensions}

\setcounter{equation}{0}

\def\theequation{\thesection.\arabic{equation}}

Clarke \cite{clarke} 
has proved that any $C^{\infty}$-Riemannian manifold $V_n$
with $C^k$-Riemannian metric ($k\geq 3$) of
rank $r$ and signature $s$ can be globally $C^k$-isometrically
embedded in $E_m (p,q)$, 
where
\begin{equation}
m=p+q, \qquad p\geq n -\frac{r+s}{2} +1
\end{equation}
and
\begin{equation}
\label{compact}
q\geq \frac{n}{2} (3n+11) 
\end{equation}
if $V_n$ is compact and 
\begin{equation}
\label{noncompact}
q\geq \frac{n}{2} (2n^2+27) + \frac{5}{2} n^2 +1
\end{equation}
if $V_n$ is non--compact.  
Clearly, an analogue of this theorem is required,
where the embedding space is Ricci, rather than Riemann, flat.
Unfortunately, such a theorem does not as yet exist, but the lower 
bounds (\ref{compact}) and (\ref{noncompact}) suggest that 
more than one extra dimension may generally be needed for global 
embeddings  in Ricci--flat spaces. Nevertheless, some 
aspects of Campbell's theorem with regard to local 
and global embeddings can be highlighted by considering 
lower--dimensional examples. 

\subsection{Embedding of $(1+1)$--dimensional Spaces}
Campbell's theorem implies that 
any $(1+1)$--dimensional  space can be locally embedded in a 
3--dimensional, Ricci--flat space $M_3$. However, the Weyl tensor 
vanishes identically in three dimensions, so the embedding space 
is necessarily flat, i.e., ${^{(3)}}R_{\mu\nu\lambda\rho} \equiv 0$. 
Thus, Campbell's theorem is equivalent 
to Friedman's theorem in this case \cite{friedman}.

This would seem to imply 
that Campbell's theorem  results in a trivial
embedding of all $(1+1)$--dimensional spaces. 
It is important to emphasize, however, that the 
{\em topology} of the embedding space is not specified in this procedure, due 
to the local nature of the theorem. In principle, therefore, a 
given  space $V_n$ may be (locally) embedded into more than one  
higher--dimensional, Ricci--flat space,  each of which has a 
different global topology. This 
follows since there is usually more than one solution to 
Eqs. (\ref{metricdevelop}) and (\ref{omegadevelop}) consistent with the 
boundary conditions (\ref{init}). 
Moreover, the range of the extra coordinate $\psi$ is  not specified 
in Campbell's approach. It may be either compact or non--compact
and this will also affect the topology of the embedding space.  
Within the context of $(1+1)$ dimensions, this implies 
that the metric on $M_3$ may not cover the whole of Minkowski space. It is 
possible, therefore, that the embedding space may contain 
singularities and, indeed, it may even exhibit a non--trivial 
causal structure. 

These features may be illustrated by 
considering different embeddings of the $(1+1)$--dimensional  Minkowski space 
${^{(2)}} d\eta^2 =dt^2 -dx^2$, where $-\infty \le \{ t,x \} \le +\infty$.
It follows from Section 2 that 
one class of embedding metrics  is given by 
${^{(3)}}ds^2 = {^{(2)}} d\eta^2 -\phi^2 d\psi^2$, 
where\footnote{In the rest of this
Section Greek indices take values in the range $(0,1)$ only.}
$\partial_{\mu}\partial_{\nu} \phi =0$.
The general solution to these equations is given by $\phi 
= a_{\mu} (\psi ) x^{\mu}$, where 
$a_{\mu}$ are arbitrary functions. 

The whole of $M_3$ is covered if $\phi =1$ and $-\infty \le \psi 
\le +\infty$. However, a non--trivial embedding is given 
by the solution $\phi = x-t$ when $\psi $ is a compact coordinate. 
In this case, 
\be
\label{string1}
{^{(3)}}ds^2 = dudv -u^2d\psi^2  ,
\ee
where $u  \equiv t-x$ and $v \equiv 
t+x$ represent null coordinates  and $\psi$ is identified with $\psi +L$ 
for some arbitrary constant $L$. A linear translation on $\psi$ 
corresponds to a null boost. Since $\psi$ is compact, the geometry 
of a constant $x$ surface, with line--element $ds^2 =dt^2 - (x-t)^2 d\psi^2$, 
is given by $R \times S^1$. This surface 
resembles a lorentzian cone, because there exists a vertex at $x=t$. 
Thus, the spacetime  (\ref{string1}) is 
geodesically incomplete. It represents a lorentzian  
orbifold whose vertex moves at the speed of light 
\cite{HS1}. In 
conclusion, therefore, the topology of the embedding space is determined 
by the specific boundary conditions that are chosen when 
solving Eq. (\ref{phidevelop}), 
as well as the range of values taken by the extra coordinate. 

\subsection{Embedding of $(2+1)$--dimensional Spaces}

Further insight may be gained by considering 
the embedding of 3--dimensional spaces in four dimensions. 
As an example, we consider the line--element in $(2+1)$ dimensions
given by
\be
\label{conical}
{^{(3)}}ds^2 =dt^2 -d\rho^2 -\lambda^2 \rho^2 d\theta^2,
\ee
where $\lambda$ is a constant and $0\le \theta 
\le 2\pi$. When $\lambda =1-4m\bar{G}$, this represents 
the spacetime generated by a static point particle of mass $m$,  
where $\bar{G}$ is the gravitational constant 
in $(2+1)$ dimensions \cite{deser}.
As is well known, this space is flat for 
$\rho \ne 0$. However, it is not globally Ricci--flat. The 
non--zero components of the 
Ricci tensor  are ${^{(3)}}{R^{\rho}}_{\rho}=
{^{(3)}}{R^{\theta}}_{\theta} =2\pi 
\lambda^{-1} (\lambda -1 ) \delta^{(2)} (\rho)$, where 
$\delta^{(2)} (\rho)$ is the Dirac delta function at the origin 
of the 2--surface $t={\rm constant}$, the normalization 
of which is defined by the condition $\int_0^{2\pi} d \theta \int^{\infty}_0
d \rho \delta^{(2)} (\rho ) \rho =1$. In fact, Eq. (\ref{conical}) 
represents a conical spacetime which is flat everywhere except 
at one point corresponding to 
the vertex $\rho =0$. (The constant $t$ sections  may be viewed as 
euclidean planes in which a wedge with opening angle $2\pi (1-\lambda )$ 
is cut out and its edges identified). 

Now, following Campbell's method, the simplest embedding 
of metric (\ref{conical}) for $\rho \ne 0$ in a 
$(3+1)$--dimensional, Ricci--flat spacetime is given by 
\be
\label{cosmicstring}
{^{(4)}}ds^2 =dt^2 -d\rho^2 -\lambda^2 \rho^2 d\theta^2 -d\psi^2.
\ee
If $\psi$ is a non--compact coordinate $(-\infty \le \psi 
\le +\infty )$, Eq. (\ref{cosmicstring}) is 
the metric of a static, vacuum cosmic string,
where $\mu \equiv \bar{G}m/G$ is the linear mass density lying 
on the $\psi$--axis and $G$ is the gravitational constant in $(3+1)$ 
dimensions \cite{vilenkin,hiscock}. This
spacetime is flat for $\rho \ne 0$ 
and the surfaces defined by $t={\rm constant}$ and $\psi = {\rm constant}$ 
have the same topology as a cone. The Ricci components 
are ${^{(4)}}{R^{\rho}}_{\rho} = 
{^{(4)}}{R^{\theta}}_{\theta} =2\pi \lambda^{-1} (\lambda -1) 
\delta^{(2)} (\rho)$. 

This embedding (\ref{cosmicstring}) is 
not global, however, because the embedding space is only locally Ricci--flat.
On the other hand, 
the space defined by Eq. (\ref{conical}) 
may be globally embedded in $(3+1)$--dimensional 
Minkowski spacetime. If we define a new coordinate 
\be
z \equiv (1-\lambda^2)^{1/2} \rho   ,
\ee 
Eq. (\ref{conical}) transforms to
\be
\label{slice}
{^{(3)}}ds^2 =dt^2 -\frac{1}{1-\lambda^2} dz^2 -\frac{\lambda^2 z^2}{1
-\lambda^2} d\theta^2.
\ee
Equation (\ref{slice}) represents the induced metric on the 
$\xi =0$ hypersurface of the $(3+1)$--dimensional 
spacetime
\be
\label{hyper1}
{^{(4)}}ds^2 =dt^2 - \frac{\lambda^2}{1-\lambda^2} 
d\xi^2 -\frac{\lambda^2}{1-\lambda^2}  (\xi +z)^2 d\theta^2 
-\frac{1}{1-\lambda^2} dz^2 -\frac{2\lambda^2}{1-\lambda^2} d \xi 
dz
\ee
and the coordinate transformation 
\be
\xi =\frac{\sqrt{1-\lambda^2}}{\lambda} \rho -z 
\ee
maps this metric onto Minkowski space. 

We conclude, therefore, that the application of Campbell's theorem 
allows us to embed the spacetime 
(\ref{conical}) into (\ref{cosmicstring}) locally, but not globally, 
in the sense that all points on the spacetime (\ref{conical}) are 
included in the embedding. Since  the spacetimes represented by 
Eqs. (\ref{conical}) and (\ref{cosmicstring}) have 
a conical singularity at the point 
$\rho =0$, Campbell's theorem will only work for $\rho \ne 0$. 
The reason for this restriction is that Campbell's theorem 
assumes implicitly that the extra coordinate vector $\partial /\partial 
\xi$ is orthogonal to Eq. (\ref{conical}), as can be 
seen from the general expression (\ref{D+1metric}). A global 
embedding of Eq. (\ref{conical}) in Minkowski space may be achieved, 
however,  by 
dropping this restriction. In this case, the embedding spacetime does 
not inherit the topological defect of the lower--dimensional manifold. 

\section{Discussion and Conclusions}

\setcounter{equation}{0}

\def\theequation{\thesection.\arabic{equation}}

We have employed Campbell's embedding theorem 
in a number of settings. Firstly, we considered spacetimes for
which the Killing vector is covariantly constant. This class includes
a number of physically interesting spaces, such as the electromagnetic
and gravitational plane waves, as well as the more general
plane--fronted waves.  
Although we found embedding spaces for waves with arbitrary amplitude,
these embeddings could in principle be generalized by finding new
solutions to Eqs. (\ref{metricdevelop}) and (\ref{omegadevelop}). We
also considered spacetimes in which the Killing vector is not
covariantly constant, including those which are solutions to vacuum GR
such as the stationary van Stockum solutions. An embedding for the
general class of $n$--dimensional Einstein spaces was found and we
also discussed the local and global embedding of some
lower--dimensional spaces. 

Campbell's theorem 
is closely related to Wesson's interpretation of 5--dimensional, 
vacuum Einstein gravity \cite{wesson,wesson1,wesson2}. 
In view of this, it would be of interest to 
consider the embedding of 4--dimensional, cosmological 
solutions in 5--dimensional, Ricci--flat spaces. For 
example, inflationary cosmology is thought to be relevant to 
the physics of the very early Universe \cite{inflation,review}. 
During inflation, the 
scale factor of the Universe accelerates and this 
acceleration is driven by the potential energy associated with 
the self--interactions of a scalar field. Different inflationary 
solutions correspond to different 
functional forms for the potential of this field. However, Campbell's theorem 
implies that all inflationary solutions can also be generated, 
at least in principle, from 
5--dimensional, vacuum Einstein gravity. This 
implies the existence of a 
correspondence between inflationary cosmology
and Einstein's theory in five dimensions. In principle, such a relationship 
could be formulated  by employing Campbell's theorem. 

Although Campbell's theorem relates 
$n$--dimensional theories to  vacuum $(n+1)$--dimensional theories, 
it does not establish a strict equivalence between them. It is therefore 
important to determine  when such theories are {\em equivalent}. Clearly, 
this is a more severe restriction than 
embedability. Two notions of equivalence that could be 
considered are {\em dynamical equivalence} and {\em 
geodesic equivalence}. Dynamical equivalence would imply that
the dynamics of vacuum $n$-dimensional theories
is included in the vacuum $(n+1)$-dimensional theories.
In that case, the embedding would be  given by Eq. (\ref{D+1metric}) 
with $\phi =1$. This would then imply that  
${^{(n+1)}}R_{\alpha \beta }= {^{(n)}}R_{\alpha \beta } = 0$ and that 
${^{(n+1)}}R_{\alpha n} = {^{(n+1)}}R_{nn} =0$ \cite{rtz}.

Alternatively, one may consider geodesic equivalence, 
in the sense of Mashhoon {\em et al.} \cite{mashhoon94}. 
In this case the
$(3+1)$ geodesic equation induces a
$(2+1)$ geodesic equation plus a force term $F^{\alpha}$:
\be
\frac{d^2 x^{\alpha}}{ds^2} 
+ {\Gamma^{\alpha}}_{\beta \gamma} \frac{dx^{\beta}}{ds}
\frac{dx^{\gamma}}{ds} = F^{\alpha}.
\ee
For geodesic equivalence one would therefore require
$ F^{\alpha} =0$, 
which is clearly so when 
\be
\frac{\partial g_{\alpha\beta}}{\partial \psi} =0.
\ee 
It would be interesting to
ask whether these equivalences hold in more general settings.
 
\vspace{1cm}

{\bf Acknowledgments} JEL was supported by the Particle Physics 
and Astronomy Research Council (PPARC), UK. CR was 
supported by CNPq (Brazil).
RT benefited from SERC UK Grant No. H09454.
SR was supported by a PPARC studentship.
CR thanks the School of Mathematical 
Sciences for hospitality where part of this work was carried out. 
We thank Roustam Zalaletdinov for discussions.
We would also like to thank the referees for helpful comments.


\end{document}